\documentclass[aps,12pt,superscriptaddress,amsfonts,amssymb,amsmath]{revtex4-2}
\pdfoutput=1

\usepackage{graphicx}
\usepackage{epsfig}
\usepackage{makeidx}
\usepackage{epstopdf}
\usepackage{xcolor}
\usepackage{amsmath}
\usepackage{bm}
\usepackage{siunitx}
\usepackage{float}
\usepackage{booktabs}  
\usepackage{array}  
\usepackage{verbatim}
\usepackage{subfig}
\usepackage{braket}
\usepackage{appendix}
\usepackage{leftidx}
\usepackage{empheq}
\usepackage{dsfont}
\usepackage{caption}

\newcommand{\angstr}{\text{\AA}}

\begin{document}

\title{Spectral Analysis of Proton Eigenfunctions \\ in Crystalline Environments}

\author{L.\ Gamberale \footnote{Email address: luca.gamberale@unimib.it}}
\affiliation{Quantumatter Inc., Dover DE USA\\ LEDA srl, Università Milano Bicocca \\ I-20126 Milano, Italy}

\author{G.\ Modanese \footnote{Email address: giovanni.modanese@unibz.it}}
\affiliation{Free University of Bozen-Bolzano \\ Faculty of Engineering \\ I-39100 Bolzano, Italy}
\date{\today}

\linespread{0.9}

\begin{abstract}
The Schr\"odinger equation and Bloch theorem are applied to examine a system of protons confined within a periodic potential, accounting for deviations from ideal harmonic behavior due to real-world conditions like truncated and non-quadratic potentials, in both one-dimensional and three-dimensional scenarios. Numerical computation of the energy spectrum of bound eigenfunctions in both cases reveals intriguing structures, including bound states with degeneracy matching the site number $N_w$, reminiscent of a finite harmonic oscillator spectrum. In contrast to electronic energy bands, the proton system displays a greater number of possible bound states due to the significant mass of protons. Extending previous research, this study rigorously determines the constraints on energy gap and oscillation amplitude of the previously identified coherent states. The deviations in energy level spacing identified in the computed spectrum, leading to minor splitting of electromagnetic modes, are analyzed and found not to hinder the onset of coherence. Finally, a more precise value of the energy gap is determined for the proton coherent states, ensuring their stability against thermal decoherence up to the melting temperature of the hosting metal.
\end{abstract}

\maketitle

\section{Introduction}

\textcolor{black}{
Over the last decades, extensive theoretical research and the development of new experimental techniques have made possible the manipulation of coherent and collective phenomena in quantum optics at the nano-scale level, such as superradiance and coherent population trapping \cite{brandes2005coherent}. These phenomena often involve quantum phase transitions \cite{heyl2018dynamical} and an ultra-strong coupling regime of light/matter interaction \cite{forn2019ultrastrong}. Their theoretical description must take into account dissipation, as is the case in general for open quantum systems \cite{bergman2003surface,stockman2010spaser,bordo2017cooperative,link2020dynamical,azzam2020ten}, which are becoming increasingly relevant for applications to nano-electronics \cite{weinbub2022computational}.}

\textcolor{black}{
Superradiance and the Dicke effect occur when an ensemble of molecules or quantum oscillators confined within a sub-wavelength region collaboratively emit and absorb coherent radiation. Coherence among these emitters is mediated by the electric radiation field. In some cases, the field-matter coupling can be significantly enhanced by the presence of surface plasmons, resulting in what is known as the plasmonic Dicke effect \cite{dicke1954coherence,andreev1980collective,sivasubramanian2001gauge,sivasubramanian2001super,pustovit2009cooperative,pustovit2010plasmon,scheibner2007superradiance,temnov2005superradiance,greenberg2012steady}.}

In this work we continue our investigation of an idealized system comprising $N_s$ charges oscillating within a lattice structure, exhibiting a superradiant transition above a specific threshold density. The dynamics of this system involve bulk plasmons and enhanced field-matter coupling, topics that we studied analytically in \cite{gamberale2023coherent}, where we demonstrated in the large $N_s$ limit the existence of an energy gap for a specific set of wavefunctions displaying coherence in both the matter and field sectors.

The oscillating charges that we have considered in our model are protons, with reference to the concrete physical case of hydrogen loading in metals, where protons are bounded to tetrahedral or octahedral sites. The large mass of protons, compared to electrons, simplifies the treatment because it allows to disregard the fermionic character of their wavefunctions, at least in certain conditions. It also turns out that the large mass of protons plays a crucial role in determining the energy spectrum and the spatial extension of their bounded wavefunctions. Here lies the major new contribution of the present work towards an improvement of the model in Ref.\ \cite{gamberale2023coherent}. In \cite{gamberale2023coherent} we have described the bound states of protons as those of ideal harmonic oscillators, introducing an ad-hoc physical cutoff on their oscillation amplitudes. Here we provide a solid theoretical justification to the existence of the limitation in the oscillation amplitude and compute numerically the exact states of the protons in a crystal structure with periodic but finite potential wells. As could be intuitively anticipated, the proton wavefunctions are much more localized than those of electrons. Furthermore, the count of bound excited states is precisely determined by the crystal structure. For instance, in the case of octahedral voids with parameters as detailed in Table \ref{tbl:Parameters}, there are precisely 12 bound states. Only these states actively contribute to the superradiance phenomenon. This elucidates and justifies the previously mentioned cutoff on oscillation amplitude.

In \cite{gamberale2023numerical}, we recall the Hamiltonian of our model and the canonical transformations of the photon field operators used to diagonalize the photonic term. That work succinctly summarizes key findings from \cite{gamberale2023coherent} in a self-consistent manner, and presents numerical simulations showing the occurrence of a quantum phase transition for small values of $N_s$, also in the presence of dissipation.

\textcolor{black}{
The final Hamiltonian in the dipole approximation takes the form:
\begin{equation}
    H_{\text{tot}}=H_{\text{osc}}+\omega'\sum_{i=1}^3 \left( C^\dagger_i C_i + \frac{1}{2} \right) + \frac{i\omega_p}{2\sqrt{{N_s}}} \sqrt{\frac{3}{8\pi}} \sum_{n=1}^{N_s} \sum_{i=1}^3 \left[ a^\dagger_{n,i}C_i - a_{n,i}C^\dagger_i 
    + a^\dagger_{n,i}C^\dagger_i - a_{n,i}C_i\right].
\label{H-final}
\end{equation}
Here, $C_i$ and $C^\dagger_i$ represent linear combinations of photon annihilation and creation operators, obtained by projecting along the three spatial directions $\mathbf{\hat{e}}_i$ and summing over all possible momentum directions $\mathbf{k}$. Using these operators, the vector potential simplifies to $\mathbf{A}=\frac{1}{\sqrt{2\omega' V}} \sum_{i=1}^3 (C_i+C^\dagger_i) \mathbf{\hat{e}}_i$.
The term $H_{\text{osc}}$ in (\ref{H-final}) denotes the sum of the Hamiltonians of the proton oscillators: $
H_{\text{osc}}=\omega' \sum_{n=1}^{N_s} \left[ \mathbf{a}^\dagger_n(t) \mathbf{a}_n(t) + \frac{3}{2} \right]$, where the $\mathbf{a}_n$ are standard destruction operators for the harmonic oscillators.
For the definition of the frequencies $\omega$, $\omega_p$, $\omega'$, please refer below. We utilize units in which $\hbar=c=1$.}

One of the key features of the model, as extensively elaborated in reference \cite{gamberale2023coherent}, is its consistent definition of the natural frequency denoted as $\omega$ for all oscillators within the material. This uniformity is established by employing a periodic electrostatic potential known as the ``jellium crystal''. During the diagonalization process of the total Hamiltonian, the frequency $\omega$ is combined with the plasma frequency $\omega_p=\sqrt{e^2N_s/(mV)}$ associated with the oscillating charges. This combination yields a modified or ``dressed'' frequency denoted as $\omega'=\sqrt{\omega^2+\omega_p^2}$. It is noteworthy that the photon momentum, represented as $\mathbf{k}$, remains unchanged throughout this process. Consequently, within this model, there exist no states in which electromagnetic energy generated within the material can propagate into the vacuum, thereby forming a naturally occurring resonating cavity.

Another crucial aspect of the emerging coherent electromagnetic modes, which oscillate with a fixed phase relation to the matter oscillators, is the orientation of their wave vectors in all possible spatial directions. This enhances the coupling between matter and electromagnetic fields, enabling the formation of the energy gap.

\textcolor{black}{
While the overall framework seems largely satisfactory in its primary aspects, a notable question remains regarding size of the oscillation amplitude of the coherent state, a parameter directly influencing the energy gap's magnitude. In this investigation, we approach this issue employing a rigorous physical methodology, thereby dispelling any lingering remnants of heuristic speculation. This is the main result of the present work.
}

In the subsequent sections, we will focus on the bound states of protons bound to crystal sites. In Sect.\ \ref{sec:core} we find the spectrum through a numerical solution of the Schr\"odinger equation and taking into account the Bloch theorem. In Sect.\ \ref{quasi} we consider a perturbation term which accounts for deviations in energy level spacing, resulting in minor e.m.\ mode splitting. 

Finally, we rigorously calculate the maximum amplitude of the coherent oscillation, which had been manually set using a heuristic argument in \cite{gamberale2023coherent}, and summarize our findings.

\section{Single-Particle Bound States in a Finite Quadratic Periodic Potential\label{sec:core}}

In reference \cite{gamberale2023coherent}, we tackled the issue of one-particle bound states within an ideal periodic infinite potential well. In idealized scenarios, such wells provide a simplified framework for understanding quantum systems. However, when considering real-world conditions, oscillators, or particles, deviate from ideality due to truncated and non-quadratic potentials. This departure from ideal behavior carries significant consequences for the structure of bound energy levels. In the subsequent analysis, we will delve into the profound implications of these deviations, examining how they impact the characteristics of bound energy states and, more importantly, its consequences in the properties of their coherent states.

\begin{figure}[H] 
   \includegraphics[width=0.8\columnwidth]{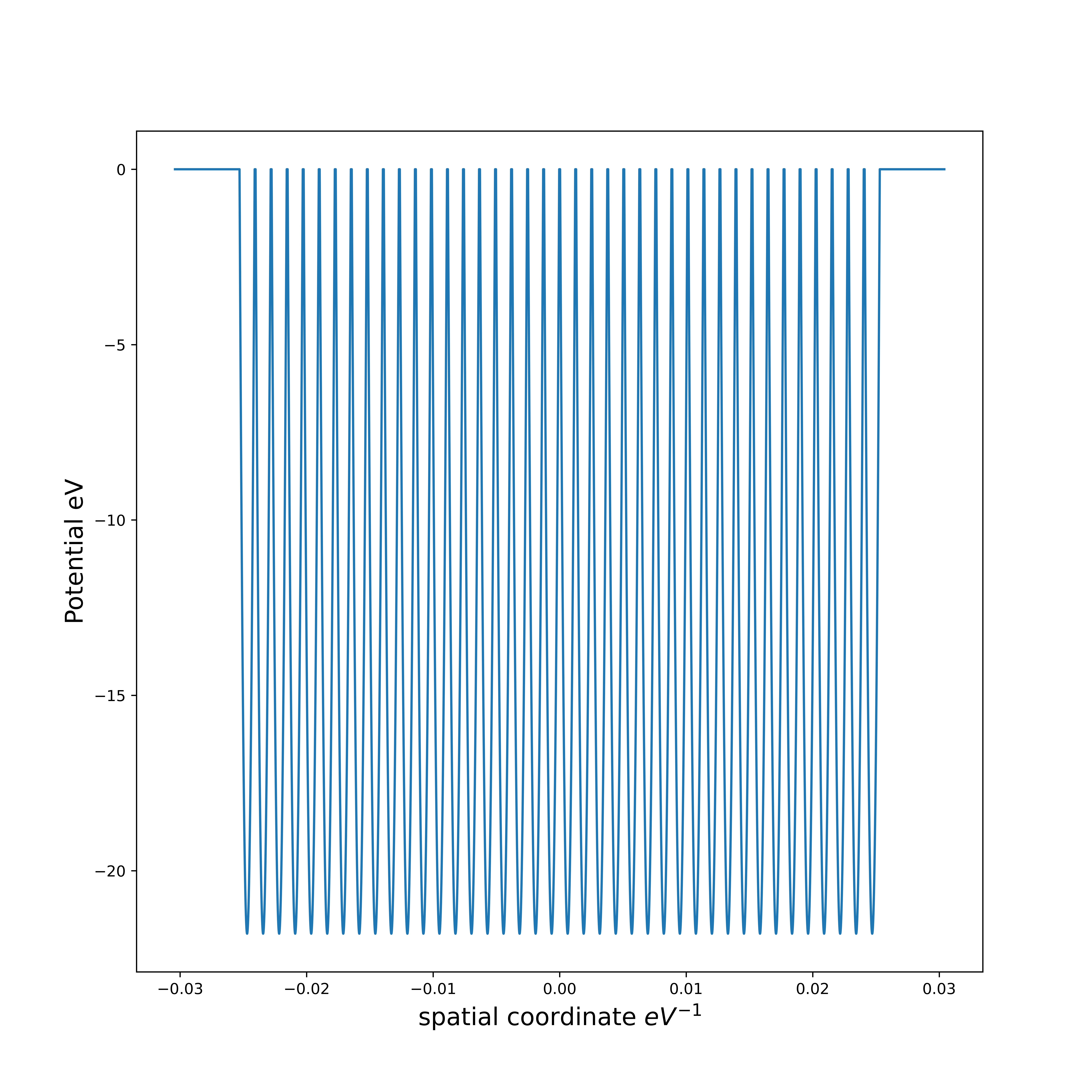}
        \caption{
                \label{fig:periodicpotential} 
                Periodic 1D potential with 40 sites with lattice spacing $d=\frac{a}{\sqrt{2}}=2.5\cdot10^{-8}\text{ cm}=1.27\cdot10^{-3}\text{ eV}^{-1}$ along the base vector $\vec a_1$.
        }
\end{figure}

Consider an ensemble of protons with mass $m$ immersed in a periodic \textcolor{black}{quasi-harmonic truncated} potential in 3D of the form
\begin{equation}
V(\vec x) = \omega\sum_{cijk}\mathcal V(\sqrt{m\omega}|\vec x - \vec x_{cijk}|)
\label{eq:sumpot}
\end{equation}
where $\omega$ is the free vibration frequency, $\vec x_{cijk}$ are the 3D positions of the lattice sites and 
\begin{equation}
\mathcal V(q) = \frac12(q^2 - q_{max}^2)\theta( q_{max}^2 -q^2)
\end{equation}
where $q_{max}$ is a parameter that controls the width and depth of the potential well \textcolor{black}{(the argument $q$ of $\mathcal V(q)$ is a mute variable)}. \textcolor{black}{The coordinates $\vec x_{cijk}$ are selected to represent various periodic crystal structures. We opt to utilize the face-centered cubic (FCC) symmetry, characterized by
\begin{equation}
    x_{cijk}=\vec a_c+(i, j, k)a
\end{equation}
where $i, j, k \in \mathbb{Z}$, $c=0, 1, 2, 3$,  $\vec a_0=(0, 0, 0)$, $\vec a_1=(0, \frac12, \frac12)$, $\vec a_2=(\frac12, 0, \frac12)$, $\vec a_3=(\frac12, \frac12, 0)$, and $a$ denotes the crystal spacing ($a=3.52\angstr$ for Nickel). The Heaviside (step) function $\theta$ ensures the potential remains non-positive, as depicted in Fig. \ref{fig:periodicpotential} along the $x$-direction with $y=z=0$.
}
The Schr\"odinger equation that determines the eigenstates of the single-particle problem is 
\begin{equation}
\left[-\frac{1}{2m}\nabla^2+V(\vec x)\right]\psi(\vec x)=E\psi(\vec x)
\label{eq:3Dshreoedinger}
\end{equation}
whose solution can be afforded using the Bloch theorem. By setting
\begin{equation}
\psi_{\vec K,\vec n}(\vec x)=e^{i\vec K\cdot\vec x}u_{\vec n}(\vec x)
\label{eq:blochfun}
\end{equation}
and working in the tight binding approximation \cite{slater1954simplified}  we can set $u_{\vec n}(\vec x)=\sum_{ijk}\phi_{\vec n}(\vec x-x_{ijk})$ where $\phi_{\vec n}(\vec x)$ are the eigenfunctions of the Schr\"odinger equation for the isolated potential $ \omega\mathcal V(\sqrt{m\omega}|\vec x|)$. 

The validity of this approximation stems from the observation that the particles under examination have a mass much larger than that of the electrons, resulting in their wave-functions displaying significant localization around the centers of the potential wells. This localization, in turn, guarantees that the level of overlap between the wave-functions associated with neighboring sites can be considered practically negligible to a considerable degree at least up to a certain level of excitation of the bound states.

The energy spectrum of the bound eigenfunctions described by Equation \eqref{eq:3Dshreoedinger} can, in fact, be derived from the spectrum of the equation
\begin{equation}
\left[-\frac{1}{2m}\nabla^2+\omega\mathcal V(\sqrt{m\omega}|\vec x|)\right]\phi_{\vec n}(\vec x)=E_{\vec n}\phi_{\vec n}(\vec x).
\label{eq:3Dshreoedinger1}
\end{equation}
independent of the momentum vector $\vec K$, that can take any value.

To substantiate the above observations, we now study numerically the 1D version of the problem, described by the  Schr\"oedinger equation
\begin{equation}
\left[-\frac{1}{2m}\frac{\partial^2}{\partial x^2}+V(x)\right]\psi(x)=E\psi(x)
\label{eq:1Dshreoedinger}
\end{equation}
where the 1D potential is given by ($\tilde x_i=d(i+\frac12)$, $-(N_w-1)/2\le i<(N_w-1)/2$, where $N_w$ is the even number of wells considered in the calculation)
\begin{equation}
V(x) = \omega\sum_{i}\mathcal V\left(\sqrt{m\omega}(x-\tilde x_i)\right).
\label{eq:singlepot}
\end{equation}

Through a rotational symmetry argument, it can be demonstrated that the 1D problem is analogous to the 3D problem and shares the same energy spectrum. Specifically, it can be shown that Eq. \eqref{eq:1Dshreoedinger} corresponds to the radial component of the Schrödinger equation in three dimensions (3D) for a single potential well (Eq. \eqref{eq:3Dshreoedinger1}), focusing on the reduced radial solutions characterized by zero angular momentum ("$s$-wave" solutions). Consequently, the numerical results we aim to present remain valid even within the framework of the complete 3D scenario.

By setting the discretization points $x_j=\varepsilon j$, $-(M-1)/2\le j\le (M-1)/2$ \textcolor{black}{($M$ arbitrary odd integer)} where the discretization step is $\varepsilon=\frac{d}{N_c}$ where $N_c$ is the number of discretization points per cell,
Eq. \eqref{eq:1Dshreoedinger} is rewritten in discretized form as
\begin{equation}
\left[-\mathbf T+2m\varepsilon^2\mathbf V\right]\bm{\psi}=2m\varepsilon^2 E \bm{\psi}
\label{eq:1Dshreoedinger2}
\end{equation}
where $\mathbf T$ is the second derivative operator in discretized form and is defined by $\mathbf T_{i,i\pm 1}=1$, $\mathbf T_{i,i}=-2$ for $-M/2\le i\le M/2$,  $T_{i,j}=0$ in all other cases, $\mathbf V_{ij}=V(x_i)\delta_{ij}$ \textcolor{black}{and $\bm\psi$ is a vector whose components are the values of $\psi$ at the points $x_i$:} $\bm\psi_i=\psi(x_i)$.

The final parameter requiring specification is the selection of the value for $q_{max}$. In the context of addressing the challenge of hydrogen loading in metals within the octahedral phase, prior literature indicates that the available radius within the octahedral voids is approximately 0.41 $d/2$. Thus, our decision is to set $q_{max}=0.41\sqrt{m\omega}\frac{d}{2}$.

\begin{figure}[H] 
       \includegraphics[width=0.8\columnwidth]{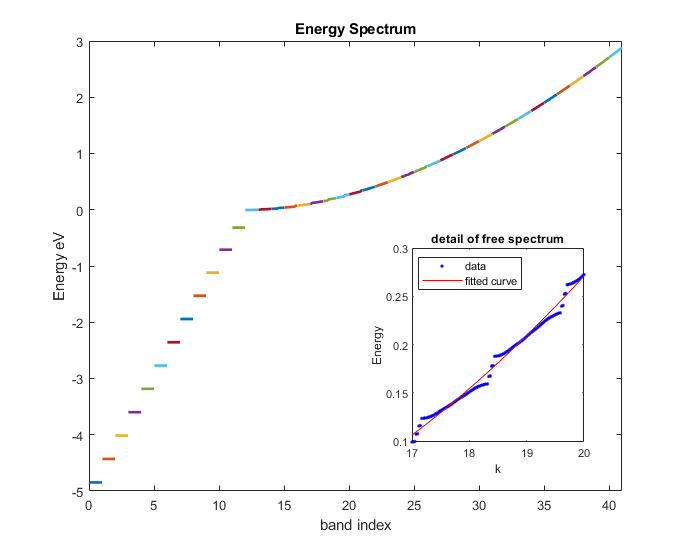}

        \caption{
                \label{fig:Energy_spectrum_periodic_potential} 
                One particle energy spectrum calculated using the parameters of Table \ref{tbl:Parameters}. Note the harmonic-oscillator-like energy spacing of the bound states and their degeneracy, equal to the number of lattice sites $N_w$. The free eigenvalues have a Brillouin-like structure. Different bans correspond to different colors. The finite number of wells causes intermediate eigenvalues to appear between adjacent bands (see sub-figure).
        }
\end{figure}

\begin{figure}[H] 
      \includegraphics[width=0.8\columnwidth]{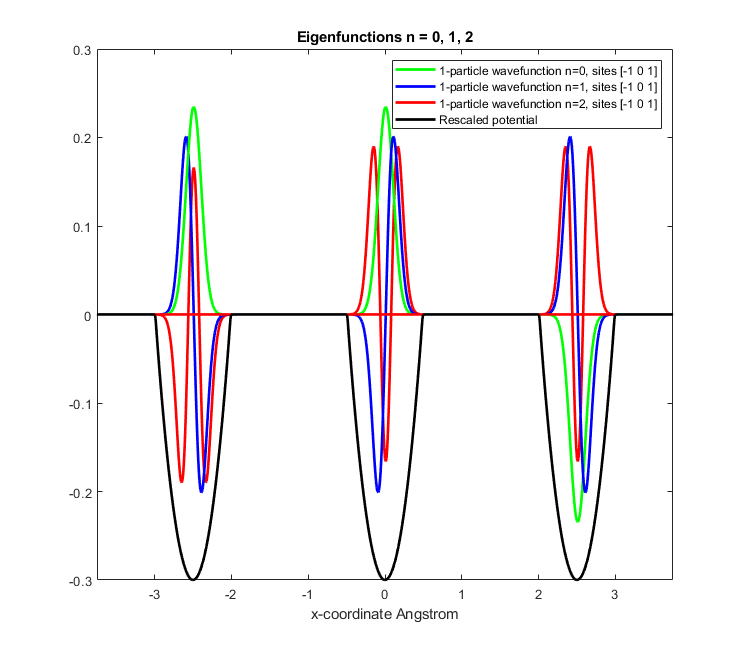}

        \caption{
                \label{fig:crystal_eigenfunctions} 
                Eigenfunctions of the bound states computed by direct diagonalization of Eq. \eqref{eq:1Dshreoedinger} with the parameters of Table \ref{tbl:Parameters}.
        }
\end{figure}

The eigenvalue problem of Eq. \eqref{eq:1Dshreoedinger2} is now completely set and can be solved numerically. The numerical solution with the parameters indicated in Tab. \ref{tbl:Parameters} is shown in Fig. \ref{fig:Energy_spectrum_periodic_potential} and reveals a very interesting structure.
The bound states with negative energy have a degeneracy equal to the site number $N_w$ as is expected for the tight binding case and their spectrum is similar to that of the harmonic oscillator. The free states (positive energy) follow a typical Brillouin dispersion relation, since the degeneracy is removed by the fact that the tight binding approximation breaks down, the wave-functions not being confined in their respective lattice sites.
The eigenfunctions of the bound states are very similar to the eigenfunctions of a harmonic oscillator centered in a specific site of the lattice (see Fig. \ref{fig:crystal_eigenfunctions}).

\begin{table}[H]
\caption{Chosen parameters for the numerical calculation}
\label{tbl:Parameters} 
\begin{tabular}{|rcl|} 
\hline
\multicolumn{1}{|c}{parameter} & \multicolumn{1}{c}{symbol} & \multicolumn{1}{c|}{value} \\
\hline
oscillation frequency & $\omega'$ & 0.41 eV\\
proton mass & $m$ & 938 MeV\\
lattice spacing & $d$ & 2.5 \angstr\\
number of points of a single cell & $N_c$ &  257\\
number of points of the potential wells & $N_w$ & 101\\
spatial discretization & $\varepsilon$ & 0.0097 \angstr\\
number of crystal sites & $N_s$ & 41\\
number of points inside the crystal & $M_c$ & 10537\\
total number of points & $M$ & 12537\\
\hline
\end{tabular}
\end{table}

One might inquire about the difference between the well-established outcomes regarding electronic energy bands, which exhibit a significantly distinct structure compared to the solutions obtained here for protons. The answer lies in the fact that, given that protons are approximately 1800 times more massive than electrons, the lower-energy wave functions are considerably more localized than those of electrons. This localization arises from the spatial scaling factor being $\frac{1}{\sqrt{m\omega}}$, consequently leading to a significantly greater number of possible bound states. In fact, the same calculation conducted for electrons reveals that the maximum number of bound states is at most 2.

\begin{figure}[H] 
      \includegraphics[width=0.8\columnwidth]{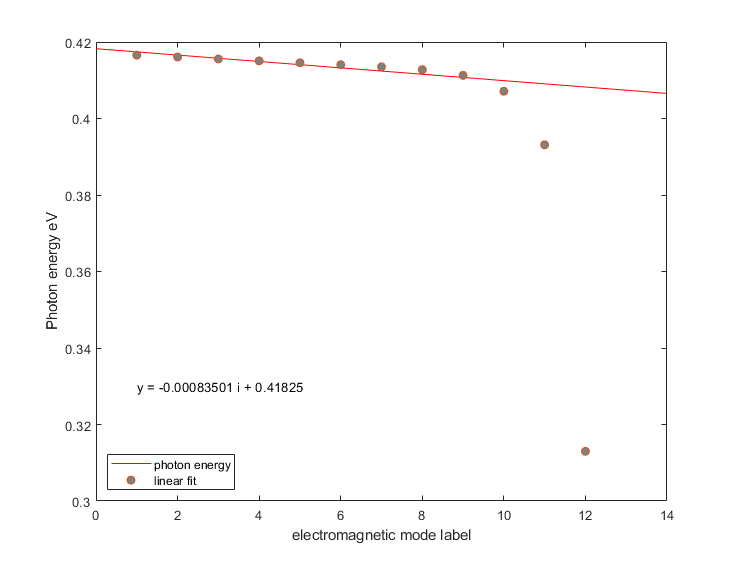}

        \caption{
                \label{fig:Photon_energy_modes} 
                Linear fit of the photon energies associated to the dipolar transitions between adjacent bound energy levels (the last two points to the right are excluded from the fit). 
        }
\end{figure}

A closer inspection reveals that the negative energy levels are not exactly equally spaced, resulting in several frequencies of the electromagnetic field being involved. Fig. \ref{fig:Photon_energy_modes} shows the dependence of the photon energies for the various couplings between adjacent energy levels. Such an issue will be dealt with in Section \ref{quasi}.

\color{black}
\section{Three-dimensional analysis}

We now extend our numerical analysis to the three-dimensional (3D) case, albeit without delving into the complexities associated with the resulting band structure for free eigenstates. Our primary objective is to demonstrate that even in the 3D scenario, bound states exhibit uniform energy spacing, a crucial factor for coherence establishment. 

Our goal is to corroborate the idea that the bound states of an isolated oscillator bound in a single potential well of the form \eqref{eq:singlepot} has an energy spectrum practically identical to that of a 3D harmonic oscillator with the same mass and oscillation frequency confined by a full harmonic potential (with no truncation) with minimum potential energy $-\frac12m\omega^2R_w$.

Equation \eqref{eq:3Dshreoedinger1} describes a single particle interacting with a spherically symmetric potential well centered at the origin of the coordinates. The potential \( \mathcal V \), always non-positive, delineates a finite set of bound states with negative energy and a continuous set of improper eigenfunctions with positive energy. However, in the spectral sector of our interest, the eigenfunctions of the harmonic oscillator decay exponentially at large distances, becoming negligible even at a distance from the origin of less than $\frac{d}{2}$.

We thus develop the spectral decomposition of Eq. \eqref{eq:3Dshreoedinger1} based on the eigenfunctions of the 3D harmonic oscillator with an unbounded potential centered at the origin \( V^h(r) = \frac{1}{2}m\omega^2(r^2 - R_w^2) \) with \( R_w=(m\omega)^{-1/2}q_{\text{max}} \). Consequently, the eigenfunctions and eigenvalues of the harmonic oscillator with negative energies closely resemble those of Equation \eqref{eq:3Dshreoedinger1} with a high degree of accuracy.

The eigenfunctions of the 3D harmonic oscillator are given by
\begin{equation}
    \phi^h_{\vec n_rlm}(\vec x)=R^h_{n_rl}(|\vec x|)Y_{lm}(\hat x),
\end{equation}
and the corresponding eigenvalues are
\begin{equation}
    E_n=(n+\frac32)\omega-\frac12m\omega^2R_w^2,
    \label{eq:theorlevels}
\end{equation}
where \( n=2n_r+l \) is the principal quantum number, \( n_r \) is the radial quantum number, and \( l \) and \( m \) are the angular quantum numbers. Here, \( Y_{lm}(\hat x) \) represents the spherical harmonics, and \( R^h_{n_rl}(r) \) denotes the radial eigenfunctions of the 3D harmonic oscillator. These radial eigenfunctions are given by
\begin{equation}
    R^h_{n_rl}(r)=
    N_{n_rl}\rho^le^{-\frac12\rho^2}L_{n_r}^{l+\frac12}(\rho^2)
    \label{eq:eigenosc3Dinf}
\end{equation}
where \( \rho=\sqrt{m\omega}r \), \( L_{n_r}^{l+\frac12} \) are the generalized Laguerre polynomials, and \( N_{n_rl} = \sqrt{\frac{m\omega2^{l+2}(2n_r)!! }{\sqrt{\pi} (2l+2n_r +1)!!}} \) are normalization factors.

To validate our approach, we numerically solved the discretized eigenvalue equation for a radial potential with truncated harmonic behavior at radius \( R_w \). Utilizing a MATLAB algorithm, we discretized the eigenvalue equation using the finite element method on a uniform lattice. The discretization method follows the same line of the 1D case, aside from the different choice of the coordinate range and the adoption of the reduced wave-function $\phi(r)\rightarrow\frac{\phi(r)}{r}$.

The radial potential considered comprises a truncated harmonic potential plus the centrifugal term and is given by
\begin{equation}
    V(r_i) = \frac{1}{2} m \omega^2 (r_i^2 - R_w^2) \cdot \theta(r_i^2 - R_w^2) + \frac{l(l+1)}{2mr_i^2}
\end{equation}
where \( r_i=\delta+i\cdot \varepsilon \) for \( i=0, 1,... N_\infty \) and \( \varepsilon=R_\infty/N_\infty \). Here, \( N_\infty \) defines the dimensionality of the eigenvalue problem and is chosen to be much larger than the number of bound eigenvalues of interest. The parameter \( \delta \) is a small number compared to \( \varepsilon \) and prevents numerical overflow when evaluating the potential at \( r=0 \).

We solved the system for a range of principal quantum numbers \( n \) and angular quantum numbers \( l \), with a maximum value of \( n_{\text{max}} = 12 \) chosen as a cutoff parameter.

We observed that the system's eigenvalues are influenced solely by the principal quantum number \( n \) with no dependence on the angular quantum number \( l \), thus reproducing the expected analytical result. Remarkably, this property remains valid even for eigenvalues very close to zero, where the influence of the non-harmonicity of the potential may be significant.

\begin{figure}[H] 
      \includegraphics[width=0.8\columnwidth]{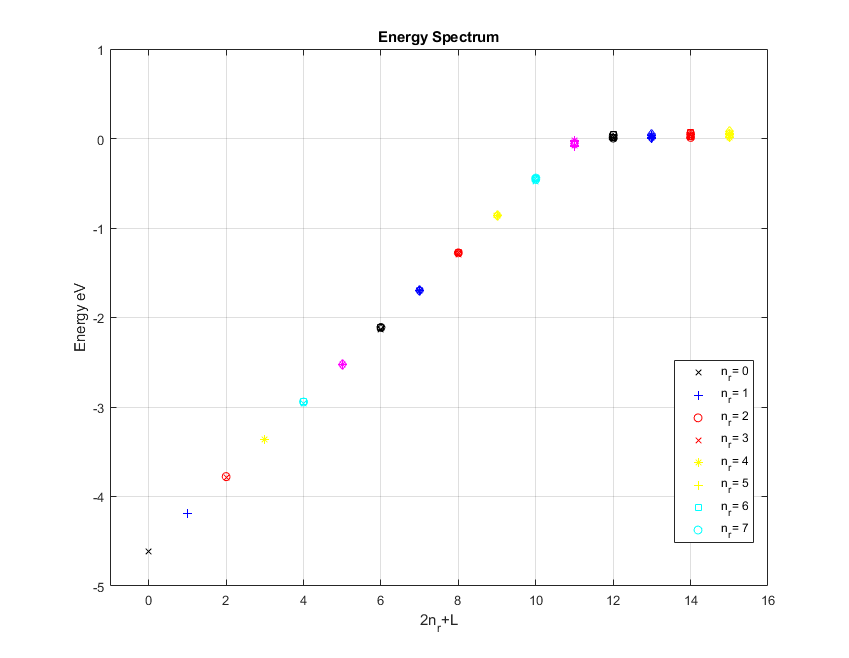}

        \caption{
                \label{fig:spectrum3Dradial} 
Eigenvalues of the radial equation for all quantum numbers of the bound states. The confirmation of independence from angular momentum, as displayed by the eigenstates of the full harmonic oscillator, persists even for eigenvalues very close to the free spectrum.
        }
\end{figure}

\begin{table}[htbp]
\centering
\caption{Calculated and theoretical energies for bound states with $N_\infty=4096$, $R_\infty= 7.5 \angstr$, $R_w = 0.5125 \angstr$
}
\label{tbl:energies}
\begin{tabular}{cccccc}
\toprule
    &      &     & Computed& Theoretical\\
$n$ & $nr$ & $l$ & Energy (eV) & Energy (eV) (Eq. \eqref{eq:theorlevels}\\
\midrule
0 & 0 & 0 & \(-4.876\) & \(-4.877\) \\
1 & 0 & 1 & \(-4.460\) & \(-4.460\) \\
2 & 1 & 0 & \(-4.042\) & \(-4.043\) \\
2 & 0 & 2 & \(-4.043\) & \(-4.043\) \\
3 & 1 & 1 & \(-3.626\) & \(-3.626\) \\
3 & 0 & 3 & \(-3.626\) & \(-3.626\) \\
4 & 2 & 0 & \(-3.208\) & \(-3.209\) \\
4 & 1 & 2 & \(-3.209\) & \(-3.209\) \\
4 & 0 & 4 & \(-3.209\) & \(-3.209\) \\
5 & 2 & 1 & \(-2.792\) & \(-2.792\) \\
5 & 1 & 3 & \(-2.792\) & \(-2.792\) \\
5 & 0 & 5 & \(-2.792\) & \(-2.792\) \\
6 & 3 & 0 & \(-2.374\) & \(-2.375\) \\
6 & 2 & 2 & \(-2.375\) & \(-2.375\) \\
6 & 1 & 4 & \(-2.375\) & \(-2.375\) \\
6 & 0 & 6 & \(-2.375\) & \(-2.375\) \\
7 & 3 & 1 & \(-1.958\) & \(-1.958\) \\
7 & 2 & 3 & \(-1.958\) & \(-1.958\) \\
7 & 1 & 5 & \(-1.958\) & \(-1.958\) \\
7 & 0 & 7 & \(-1.958\) & \(-1.958\) \\
8 & 4 & 0 & \(-1.540\) & \(-1.541\) \\
8 & 3 & 2 & \(-1.541\) & \(-1.541\) \\
8 & 2 & 4 & \(-1.541\) & \(-1.541\) \\
8 & 1 & 6 & \(-1.541\) & \(-1.541\) \\
8 & 0 & 8 & \(-1.541\) & \(-1.541\) \\
9 & 4 & 1 & \(-1.126\) & \(-1.124\) \\
9 & 3 & 3 & \(-1.125\) & \(-1.124\) \\
9 & 2 & 5 & \(-1.125\) & \(-1.124\) \\
9 & 1 & 7 & \(-1.124\) & \(-1.124\) \\
9 & 0 & 9 & \(-1.124\) & \(-1.124\) \\
10 & 5 & 0 & \(-0.713\) & \(-0.706\) \\
10 & 4 & 2 & \(-0.713\) & \(-0.706\) \\
10 & 3 & 4 & \(-0.711\) & \(-0.706\) \\
10 & 2 & 6 & \(-0.709\) & \(-0.706\) \\
10 & 1 & 8 & \(-0.707\) & \(-0.706\) \\
10 & 0 & 10 & \(-0.707\) & \(-0.706\) \\
11 & 5 & 1 & \(-0.316\) & \(-0.289\) \\
11 & 4 & 3 & \(-0.310\) & \(-0.289\) \\
11 & 3 & 5 & \(-0.303\) & \(-0.289\) \\
11 & 2 & 7 & \(-0.296\) & \(-0.289\) \\
11 & 1 & 9 & \(-0.292\) & \(-0.289\) \\
11 & 0 & 11 & \(-0.290\) & \(-0.289\) \\
12 & 6 & 0 & \(-0.002\) & \(0.128\) \\
12 & 5 & 2 & \(0.001\) & \(0.128\) \\
12 & 4 & 4 & \(0.002\) & \(0.128\) \\
12 & 3 & 6 & \(0.004\) & \(0.128\) \\
12 & 2 & 8 & \(0.006\) & \(0.128\) \\
12 & 1 & 10 & \(0.008\) & \(0.128\) \\
12 & 0 & 12 & \(0.011\) & \(0.128\) \\
\bottomrule
\end{tabular}
\end{table}
In Table \ref{tbl:energies} are shown the computed eigenvalues for the bound states together with the corresponding theoretical values for the 3D harmonic oscillator and the results are found to be in excellent agreement with theoretical predictions. Figure \ref{fig:spectrum3Dradial} illustrates the dependence of the calculated energies on \( n \) and \( l \), clearly highlighting the uniformity of the level spacing for bound states only.

\color{black}

\section{Quasi-Harmonic Case}
\label{quasi}

In \cite{gamberale2023coherent} we have analyzed the conditions for the onset of electromagnetic coherence for a system of $N$ charged oscillators immersed in a periodic harmonic potential and we have heuristically inferred the conditions for the non-divergence of the energy gap and oscillation amplitude. More specifically, the divergence of the cited quantities depends directly on the unlimited number of the bounded energy levels of the harmonic oscillator.

We now demonstrate that when considering the more realistic scenario in which the periodic potential is confined as outlined in Section \ref{sec:core}, both the energy gap and the oscillation amplitude are naturally constrained without any additional assumption.

Another complicating factor in our analysis arises from the non-uniform spacing of energy levels within the oscillators, leading to the coupling of several electromagnetic field modes with the oscillators, as opposed to just a single mode.

\textcolor{black}{
To tackle this intricacy, we pivot towards the simpler 1D scenario, underlining that analogous conclusions apply to the 3D counterpart. We commence by scrutinizing the idealized scenario where all oscillator levels, labeled as $E_i=(i+\frac12)\omega_0$, exhibit uniform spacing, alongside introducing a minor perturbation term of $\frac12\delta_\omega i(i-1)$ to accommodate deviations. This adjustment yields an energy disparity between successive levels of $E_{i+1}-E_i=\omega_0+\delta_\omega i$ ($\omega_0$ denotes the unperturbed frequency).
}

Commencing with $\delta_\omega=0$, we initially observe that the coherent state comprises a solitary frequency $\omega_0$. Subsequently, we incrementally increase $\delta_\omega$ by an infinitesimal amount, resulting in an exceedingly minor splitting of the various electromagnetic modes. For the sake of continuity, we anticipate that the state characterized by the minimum energy still consists of a solitary electromagnetic mode. 

Upon conducting a thorough examination of the phase transition mechanism, we ascertain that, in order to generate a negative contribution of the interaction term throughout a single oscillation period of the oscillators, 
\textcolor{black}{
it is imperative that the phase differential between the modes with frequencies $\omega_i$ originating from the $i$-th energy level and the resulting frequency $\omega'$ associated with the coherent mode remains confined to a value less than $\epsilon\pi$, where $\epsilon$ is the coupling constant of Eq. (26) in \cite{gamberale2023coherent}.
}

\textcolor{black}{
The above condition can be implemented by requiring
\begin{equation}
    |\omega_i-\omega'|<\epsilon\frac{\omega'}{2}
    \label{eq:phase_condidion}
\end{equation}
where $\omega'$ is taken as the averaged oscillation frequencies of the em field and
\begin{equation}
    \omega_i=\omega_0-|\delta_\omega |i \ \ \ \ \ \  0\le i\le i_{max}
    \label{eq:phase_condidion-2}
\end{equation}
and where $\delta_\omega$ is the coefficient of variation of $\omega_i$ with the level index (see Fig. \ref{fig:Photon_energy_modes}). 
}

\textcolor{black}{
Eq. \eqref{eq:phase_condidion-2} bears resemblance to the critical condition observed in the Kuramoto model \cite{kuramoto} for an ensemble of oscillators with marginally varying natural frequencies. The theoretical constructs delineated in \cite{gamberale2023coherent} could potentially draw parallels with the Kuramoto model, thereby furnishing a theoretical underpinning from the vantage point of quantum field theory. This correlation will be investigated in an upcoming inquiry.
}

Setting the average frequency $\omega'$ as the mean position with respect to the involved frequencies so that
\begin{equation}
    \omega'=\omega_0-|\delta_\omega|\frac{i_{max}}{2}
    \label{eq:phase_condidion-3}
\end{equation}
\textcolor{black}{
we can solve Eqs.\ \eqref{eq:phase_condidion}, \eqref{eq:phase_condidion-2}, \eqref{eq:phase_condidion-3} for the variable $i$, obtaining the maximum value as $i_{max}=\frac{2\epsilon\omega_0}{|\delta_\omega|(1+\epsilon)}$. 
Utilizing the parameters in Table \ref{tbl:Parameters} and from Appendix A in \cite{gamberale2023coherent}, we determine that the number of bound states is $N_b=i_{\text{max}}=12$ (see Fig. \ref{fig:Energy_spectrum_periodic_potential}) for the 1D case, and $N_b=i_{\text{max}}=11$ for the 3D case, one less than in the 1D case due to the zero point energy contribution (see Table \ref{tbl:energies}). Consequently, we can infer that all these bound states contribute to the establishment of coherence, resulting in an averaged frequency of $\omega'=0.4$ eV.}

Finally, we can compute the maximum amplitude of the coherent oscillation that is the parameter that ultimately controls the size of the energy gap (see Eqs. (30) and (43b) of \cite{gamberale2023coherent} for the up-to-1st and up-to-2nd order calculation). 

The number of photons within the coherent state corresponds to the available dipolar transitions in the system, which in our context amounts to $N_b-1$. Additionally, for a Glauber state, the expected number of photons is $|\alpha|^2$, thereby setting $|\alpha|^2=N_b-1$.

Utilizing Eq. (34) from \cite{gamberale2023coherent}, and given $\omega'=0.41$ eV for protons (Eq. (A24) of \cite{gamberale2023coherent}), we derive for the 3D case an estimate for the oscillation amplitude:
\begin{equation}
    f
    = 
    \frac{2}{d}\sqrt{\frac{N_b-1}{m\omega'}}\simeq 0.25,
\end{equation}
which contrasts with the heuristic estimate of 0.4 in Ref. \cite{gamberale2023coherent}, established on intuitive grounds alone. Furthermore, the energy gap value shifts from the estimated 1 eV to the more precise 0.37 eV.
\section{Conclusion}

\textcolor{black}{
In this investigation, we computed the maximum amplitude of coherent oscillation of protons in a metal hydride, constrained by the periodic electrostatic potential generated by the electron distribution within the metallic crystal. This calculation, derived from fundamental principles, plays a fundamental role in the theoretical framework describing the quantum coherence of a plasma comprising protons confined to crystal lattice sites by a quasi-harmonic potential. Moreover, this parameter dictates the ultimate energy gap attained by the coherent state.}

\textcolor{black}{
Previously, the determination of this maximum amplitude relied on heuristic considerations, guided by the understanding that protons, tethered to tetrahedral or octahedral sites within each crystal unit cell, cannot transition to adjacent cells during their oscillatory motion, particularly within the realm of rapid dynamics. However, it is acknowledged that over extended time scales, proton diffusion within the crystal lattice occurs (indeed, this process underpins the initial hydrogen loading of the system).
}

\textcolor{black}{
It is noteworthy that the energy gap obtained is significantly greater than the average thermal energy characteristic of the crystal's operational temperature range (an average thermal energy of 0.37 eV corresponds to a temperature above 4000 K). This ensures the stability of the coherent state against thermal decoherence. In essence, these coherent structures, once established within the metallic crystal, constitute a crucial element of proton dynamics and necessitate consideration in situations where collective phenomena hold paramount significance.
}

\textcolor{black}{In order to find the maximum number of quasi-harmonic energy levels involved in the coherence process, that are directly related to the coherent oscillation amplitude, we have solved the Schr\"odinger equation in the crystal (a periodic solution, according to the Bloch theorem), after reducing the full 3D problem to 1D  and defining a potential which is still harmonic but with a finite depth, and therefore has a spectrum with both bound and free states.}

\textcolor{black}{The calculated one-particle energy spectrum, employing realistic physical parameters, exhibits distinctive bands that markedly differ from electronic bands. This disparity arises from the considerable mass discrepancy between protons and electrons, coupled with the significantly more localized nature of proton wavefunctions in comparison to electrons.}

\textcolor{black}{The nonuniform spacing of energy levels, stemming from the non-harmonicity of the potential, introduces the complexity of multiple electromagnetic modes rather than a singular one. To address this challenge, we introduced a perturbation term to accommodate deviations in energy level spacing, leading to slight level splitting. Importantly, our analysis demonstrates that this splitting does not impede the onset of coherence, and all bound modes actively contribute to the process.}

To summarize, the limited number of energy levels involved in the coherent transition immediately resolves the problem related to the unlimited energy gap foreseen in Ref. \cite{gamberale2023coherent} and the related unboundedness of the oscillation amplitude. In Ref.\  \cite{gamberale2023coherent} we imposed ``by hand'' an upper bound the coherent oscillation through a heuristic physical argument. In the present paper we theoretically substantiate such an ansatz through an explicit calculation and we find a result essentially consistent with the previous result.

The energy gap identified in our analysis guarantees that the coherent proton states spontaneously formed within metal hydrides possess sufficient stability against thermal decoherence, even up to the melting temperatures of the metal itself.

\ 

\noindent
\textbf{Acknowledgment:} This work was partially supported by the Free University of Bozen-Bolzano with the research project NMCSYS-TN2815.

\ 

\noindent
\textbf{Author contributions:} Conceptualization, L.G. and G.M.;  software, L.G.; formal analysis, L.G.;  writing--original draft preparation, L.G. and G.M.; writing--review and editing, L.G. and G.M.

\bibliography{Numerical_analysis} 
\bibliographystyle{ieeetr}

\end{document}